\documentclass[twocolumn,showpacs,preprintnumbers,amsmath,amssymb]{revtex4}

\usepackage{graphicx}
\usepackage{dcolumn}
\usepackage{bm}

\begin{document}

\preprint{APS/123-QED}

\title{A New Unconventional Antiferromagnet, Yb$_{3}$Pt$_{4}$}
\author{M. C. Bennett,$^{1,2}$ P. Khalifah,$^{3,4}$ D. A. Sokolov,$^{1,3}$ W. J. Gannon,$^{1}$ Y. Yiu,$^{1,2}$ M. S. Kim,$^{1,3}$ C. Henderson,$^{5}$ and M. C. Aronson$^{1,2,3}$}

\affiliation{$^{1}$ Department of Physics, University of Michigan, Ann Arbor, MI 48109-1120}
\affiliation{$^{2}$ Department of Physics and Astronomy, Stony Brook University, Stony Brook, NY 11974}
\affiliation{$^{3}$ Brookhaven National Laboratory, Upton, NY  11973}
\affiliation{$^{4}$ Department of Chemistry, Stony Brook University, Stony Brook, NY 11974}
\affiliation{$^{5}$ Electron Microbeam Analysis Laboratory, University of Michigan, 48109-1005}

\date{\today}

\begin{abstract}
We report the synthesis and basic properties of single crystals of a
new binary compound, Yb$_{3}$Pt$_{4}$. The Yb ions in this compound
are fully trivalent, and heat capacity measurements show that the
crystal field scheme involves a doublet ground state, well separated
from the excited states, which are fully occupied above $\sim$ 150
K. The heat capacity displays a large, weakly first order anomaly at
2.4 K, where a cusp is observed in the magnetic susceptibility
signalling the onset of antiferromagnetic order. The entropy
associated with this order is the full Rln2 of the doublet ground
state, however the magnetic susceptibility in the ordered phase is
dominated by a large and temperature independent component below the
Neel temperature. The heat capacity in the ordered state originates
with ferromagnetic spin waves, giving evidence for the inherently
local moment character of the ordered state.  The electrical
resistivity is unusually large, and becomes quadratic in temperature
exactly at the Neel temperature. The absence of analogous Fermi
liquid behavior in the heat capacity and the magnetic susceptibility
implies that Yb$_{3}$Pt$_{4}$ is a low electron density system,
where the Fermi surface is further gapped by the onset of magnetic
order.

\end{abstract}

\maketitle

\section{Introduction}

Due in part to their ubiquity among different classes of strongly
correlated systems, there is longstanding interest in the effects of
quantum phase transitions on physical behavior, especially their
role in stabilizing novel electronic phases. Much experimental and
theoretical attention has focussed on magnetically ordered metals,
and the heavy electron compounds have been especially heavily
studied due to their electronic and behavioral diversity, as well as
the proven utility of the Doniach phase diagram as an organizational
scheme for the entire class of materials.~\cite{doniach} Many
magnetically ordered heavy electron compounds have been studied,
where both antiferromagnetic and ferromagnetic transitions have been
suppressed to form magnetic quantum critical
points.~\cite{stewart2001} Nonetheless, the need to make a link
between experimental results and theoretical models places
considerable constraints on experimental systems. Ideally, such
compounds should be single crystals of stoichiometric compounds,
demonstrated to have low levels of disorder. As grown, magnetic
order should occur at the lowest possible temperatures, but
continuous variables such as pressure or magnetic fields can be used
to tune the system to the quantum critical point. Tuning criticality
using compositional variation is to be avoided if at all possible,
due to the profound impact which disorder has been demonstrated to
have on some quantum critical systems. For these reasons, most
attention has focussed on a few model systems such as
YbRh$_{2}$Si$_{2}$,~\cite{gegenwart2002}
Sr$_{3}$Ru$_{2}$O$_{7}$,~\cite{grigera2004} and
CeCu$_{6-x}$Au$_{x}$.~\cite{vonlohneysen1996} It is clear that in
order to make continued progress towards exploring and refining the
available theoretical models, new quantum critical systems must be
identified which not only have diverse magnetic properties, but also
satisfy most - if not all - of these stringent criteria. We report
here the synthesis and essential properties of a new intermetallic
antiferromagnet, Yb$_{3}$Pt$_{4}$, which has great promise as a host
for quantum criticality.

\section{Experimental Details}

\begin{figure}[t]
\begin{center}
\includegraphics[width=7.0cm]{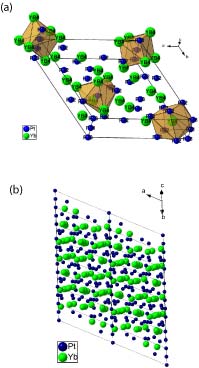}
\end{center}
\caption{\label{f1} (color online) (a) Schematic representation of the unit
cell and the polyhedra associated with each atomic site of
Yb$_{3}$Pt$_{4}$ crystallizing in the rhombohedral $hR14$ (No. 148)
structure. There are 18 Yb and 24 Pt atoms in the unit cell of Yb$_{3}$Pt$_{4}$. The Pt atoms are represented by the blue (dark gray) spheres, and the Yb atoms are represented by the green (gray) spheres, as indicated in the key.
(b) Schematic of a section of a single crystal of Yb$_3$Pt$_4$ showing
the layering of Yb and Pt atoms within the structure. Eight unit cells
are shown.}
\end{figure}

Single-crystalline samples of Yb$_{3}$Pt$_{4}$ were grown using a Pb
flux.~\cite{fisk1989} Equal molar amounts of Yb and Pt were
dissolved in molten Pb in an alumina crucible, by heating the
mixture of elements to 1200$^{\circ}$ C in vacuum in a sealed quartz
tube. The mixture was then cooled slowly to 450$^{\circ}$ C,
resulting in the formation of crystals. Molten Pb was removed by
centrifuging the crucible, isolating the single crystals, which have
typical dimensions of 0.2$\times$0.2$\times$1 mm$^{3}$. The crystals
were etched in a 50/50 by volume mixture of acetic acid and hydrogen
peroxide to remove any traces of the Pb flux remaining on the
surface. Single crystal X-ray diffraction measurements of a fragment
of one of the measured crystals determined that Yb$_{3}$Pt$_{4}$,
crystallizes in a rhombohedral $hR14$ (No. 148) space group with
lattice parameters of $a=12.8971$ \AA, $c=5.6345$ \AA, and $V =
811.65$ \AA$^{3}$, with $Z = 6$. Yb$_{3}$Pt$_{4}$ is a known
compound,~\cite{palenzona1977} but single crystals have not
previously been synthesized. Single crystals of a similar compound,
Yb$_3$Pd$_4$, have been grown and measured~\cite{politt1985,walter}
and share the same rhombohedral crystal structure and several of the
physical properties of Yb$_{3}$Pt$_{4}$, but this is the first
report of the physical properties of Yb$_{3}$Pt$_{4}$, either
polycrystalline or single crystal. The structure was solved and
refined using the SHELXL-97 program.~\cite{sheldrick} Corrections
were made for absorption and extinction, and the atomic positions
were refined with anisotropic displacement parameters. Atomic
coordinates and thermal parameters are listed in Table
\ref{tab:table1}. The primitive cell is quite simple with three Pt
positions, and just one Yb position. The full unit cell contains 18
Yb atoms, and 24 Pt atoms as illustrated in Fig.\ref{f1}a, along
with the polyhedra for each atomic position. The Pt1 and Pt3
positions have octahedral coordination, and the Pt2 and Yb4 sites
show irregular polyhedral coordination. Fig.\ref{f1}b is a schematic
of a section of a Yb$_3$Pt$_4$ single crystal consisting of 8 unit
cells, and shows a structure of alternating layers of Yb and Pt.

\begin{table}[t]
\caption{\label{tab:table1}Crystallographic data for
Yb$_{3}$Pt$_{4}$.}
\begin{tabular}{cccccc}
 Atom & Site & $x$ & $y$ & $z$
 & $U_{\rm eq}$\footnotemark[1](\AA$^{2}$)\\
\hline
Pt1 & $3a$ & 0 & 0 & 0 & 0.00754\\
Pt2 & $18f$ & 0.88391 & 0.28131 & 0.05311 & 0.00534\\
Pt3 & $3b$ & 0 & 0 & 1/2 & 0.00975\\
Yb  & $18f$ & 0.04279 & 0.21182 & 0.23494 & 0.00630\\
\hline
\end{tabular}
\footnotetext[1]{$U_{\rm eq}$ is defined as one-third of the
trace of the orthogonalized $U_{\rm ij}$ tensor.}
\end{table}

The electrical resistivity was measured in a Quantum Design Physical
Property Measurement (PPMS) System using a conventional four-probe
method between 0.4 and 300 K in zero field and in magnetic fields as
large as 9 T. Contacts were made to the crystal using silver epoxy,
which was cured at 100$^{\circ}$ C for several hours. Contact
resistances typically ranged from 1 to 10 $\Omega$. To determine the
resistance of the Yb$_3$Pt$_4$ crystal, an alternating current is
driven through the crystal at 7.5 Hz, with amplitudes ranging from
100 to 500 $\mu$A and the resulting voltage drop across the crystal
is measured. Magnetization measurements were performed using a
Quantum Design Magnetic Property Measurement System (MPMS), at
temperatures ranging from 1.8 K to 300 K, and in magnetic fields up
to 7 T. ac magnetic susceptibility measurements were also performed
in the MPMS, using a 17 Hz AC field with an amplitude of 4.17 Oe.
Specific heat measurements were performed using a Quantum Design
Physical Property Measurement System for temperatures between 0.4
and 70 K.

\section{Results and Discussion}

\begin{figure}[t]
\begin{center}
\includegraphics[width=7.0cm]{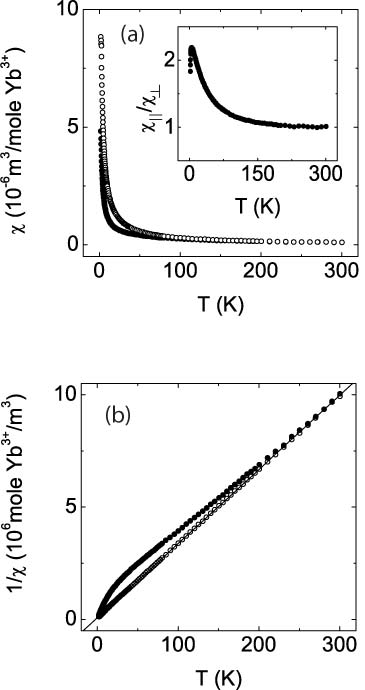}
\end{center}
\caption{\label{f2} (a) The temperature dependence of the magnetic
susceptibility of Yb$_{3}$Pt$_{4}$ with a 0.2 T field parallel
($\chi_{\parallel}$, (filled circles) and  perpendicular
($\chi_{\perp}$, hollow circles) to the c-axis. The inset shows the
magnetic anisotropy $\chi_{\parallel}$/$\chi_{\perp}$. (b) The
inverses of $\chi_{\parallel}$ (filled circles) and $\chi_{\perp}$
(hollow circles) with a Curie-Weiss fit to the data for the
perpendicular configuration (solid line).}
\end{figure}

The magnetic properties of Yb$_{3}$Pt$_{4}$ indicate that this
compound has local moment character at high temperatures. The
magnetic susceptibility measured with a field of 0.2 T parallel
($\chi_{\parallel}$) and perpendicular ($\chi_{\perp}$) to the
c-axis is plotted in Fig.\ref{f2}a. The susceptibilities increase
smoothly with decreasing temperature, and as indicated in the inset
to Fig.\ref{f2}a, the anisotropy $\chi_{\parallel}$/$\chi_{\perp}$
is near unity throughout the high temperature local moment regime,
and reaches a maximum value slightly larger than two below $\sim$
150 K.

The inverses of $\chi_{\parallel}$ and $\chi_{\perp}$ are plotted as
functions of temperature in Fig.\ref{f2}b. The same Curie-Weiss
behavior is found above $\sim$ 150 K for both field orientations,
where $\chi_{\parallel,\perp}$=C/(T+$\theta$). The effective moment
per Yb ion, deduced from the Curie constant
C=$\mu_{eff}^2$/3k$_{B}$, where the effective magnetic moment per Yb
ion, $\mu_{eff}$=g$\mu_B\sqrt{J(J+1)}$ is 4.24 $\mu_{B}$, which is
just below the Hund's rule moment for Yb$^{3+}$, 4.54 $\mu_{B}$. We
find that the Weiss temperature $\theta$=-2.3 K, indicating net
antiferromagnetic interactions among the Yb moments. Curie - Weiss
behavior is observed down to the lowest temperatures for
$\chi_{\perp}$, although 1/$\chi_{\parallel}$ is slightly enhanced
above the Curie-Weiss law for temperatures below $\sim$ 150 K. This
may result from the depopulation of the full Yb$^{3+}$ crystal field
manifold, presumed to be fully occupied for temperatures above
$\sim$ 150 K. The susceptibility data indicate that the easy axis
for the local Yb moments is perpendicular to the c-axis at low
temperatures.

\begin{figure}[t]
\begin{center}
\includegraphics[width=7.0cm]{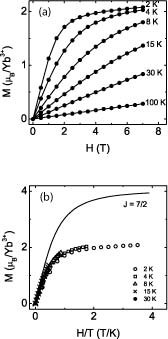}
\end{center}
\caption{\label{f3} (a) The magnetic field H dependence of the
magnetization of Yb$_{3}$Pt$_{4}$ at different temperatures T with H
perpendicular to the c-axis. (b) Data from (a) plotted as a function
of H/T. The J=7/2 Brillouin function (solid line) compares
unfavorably to the data.}
\end{figure}

Further evidence for the local moment character of Yb$_{3}$Pt$_{4}$
comes from magnetization measurements, which have been carried out
at a variety of temperatures between 2 K and 100 K (Fig.\ref{f3}a),
with the fields applied perpendicular to the c-axis.
At high temperatures, the magnetization M is linear with field H,
but a pronounced curvature develops as the temperature is reduced.
These data have been replotted in Fig.\ref{f3}b as functions of
field divided by temperature, H/T. For all temperatures, the
magnetization collapses on to a universal curve, as indicated in
Fig.\ref{f3}b, implying simple paramagnetic behavior. As we will
show below, we believe that the crystal field split manifold of
Yb$^{3+}$ states is being thermally populated above $\sim$ 30 K,
varying the effective Yb$^{3+}$ moment for temperatures below $\sim$
150 K, and indeed Fig. 3b shows that the high field magnetization
does not approach the nominal J=7/2, g=8/7 high field value of 4
$\mu_{B}$/Yb at the lowest temperatures.
\begin{figure}[t]
\begin{center}
\includegraphics[width=7.0cm]{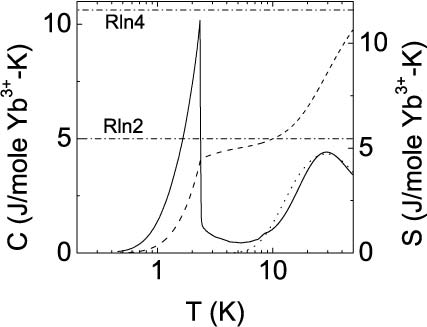}
\end{center}
\caption{\label{f4} The temperature dependence of the magnetic and
electronic heat capacity, C$_{M}$ (solid line), and its associated
entropy, S (dashed line). The dotted line indicates a fit to a
Schottky expression, yielding peaks at 26 K and 63 K.}
\end{figure}

Measurements of the heat capacity, C, show that Yb$_{3}$Pt$_{4}$
becomes magnetically ordered at low temperature. We have estimated a
phonon contribution C$_{ph}$ to the total heat capacity C by fitting
the high temperature C(T) to the Debye model, yielding a Debye
temperature $\theta_{D}$=180 K, consistent with the value
$\theta_{D}$=177 K obtained from determinations of the atomic
displacement parameters~\cite{sales2000, sales1999} taken from
single crystal x-ray diffraction measurements. The temperature
dependence of the resulting magnetic and electronic heat capacity
C$_{M}$=C-C$_{ph}$ is plotted in Fig.\ref{f4}. At low temperatures,
C$_{M}$ is dominated by a large and extremely sharp transition at
2.4 K. The entropy S calculated from these data approaches Rln2
above 2.4 K, indicating that the 2.4 K transition corresponds to the
magnetic ordering of a well-isolated doublet ground state. S/R
increases gradually from ln2 to ln4 by $\sim$ 50 K. Clear evidence
for a Schottky peak at 26 K is seen in Fig.\ref{f4}, and the
Schottky fit implies that the first and second excited states have 
energies of 50 K and 127 K. The
temperature dependence of the entropy, S, is consistent with two
crystal field split doublet states and one quartet state, as
expected for Yb in a rhombohedral local environment.~\cite{walter}

\begin{figure}[t]
\begin{center}
\includegraphics[width=7cm]{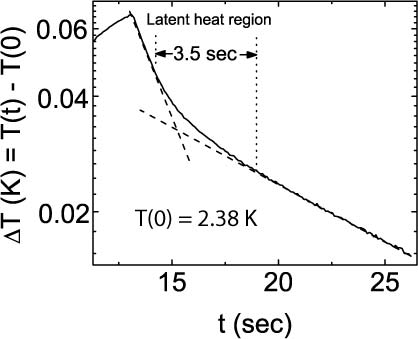}
\end{center}
\caption{\label{f5} A semi-log plot of the time dependence of the
temperature of the crystal during a heat capacity measurement near
the phase transition at 2.38 K. The non-linear region is associated
with a latent heat, as described in the text.}
\end{figure}

Fig. \ref{f4} suggests that the incipient development of critical
fluctuations is cut-off by the sudden onset of magnetic order at 2.4
K, a scenario which implies that the transition at 2.4 K is weakly
first-order and is not wholly driven by fluctuations. An analysis of
the raw specific heat data confirms that there is a latent heat
associated with the transition (Fig. \ref{f5}). The heat capacity on
each side of the transition is clearly separated into two linear
regions on the semi-log plot, the larger sloped, lower heat capacity
found above T$_{N}$, and the smaller sloped, higher heat capacity
region found below T$_{N}$. For a second order phase transition, the
two lines intersect at the critical temperature. However, for a
first order transition, the associated latent heat results in an
apparent decrease in the cooling rate near the transition
temperature. This effect is evident in Fig.\ref{f5}.

Taking known values for the thermal conductance between the crystal
and the thermal bath, the measured change in temperature during the
heat pulse, and knowing from the plot in Fig.\ref{f5} that the
undercooling effect is present for $\sim$ 3.5 seconds, we estimate
an upper bound for the latent heat to be 2$\times$10$^{-4}$ J/g, or
0.09 J/mole-Yb. Table \ref{tab:table2} is a list of the latent heats
and magnetic ordering temperatures for some rare earth elements and
rare earth based magnetic compounds that have first order
ferromagnetic
 and antiferromagnetic transitions. The latent heat of Yb$_{3}$Pt$_{4}$ is two to three orders of magnitude lower than those found in these rare earth based
 magnetic compounds, indicating that this transition is
 very weakly first order.

\begin{figure}[t]
\begin{center}
\includegraphics[width=7cm]{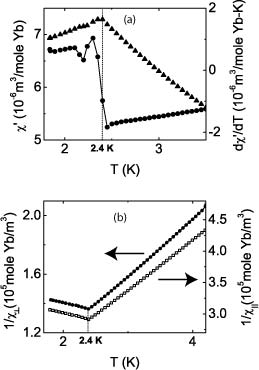}
\end{center}
\caption{\label{f6} The temperature dependence of the real part of
the zero field ac susceptibility $\chi'$, showing an
antiferromagnetic cusp and a discontinuity in its temperature
derivative, d$\chi'$/dT at the antiferromagnetic transition
temperature, 2.4 K. (b) Clear changes in slope are found for
$\chi_{\perp}^{-1}$ and $\chi_{\parallel}^{-1}$ at 2.4 K, although
Curie-Weiss behavior is found down to the Neel temperature. The
measuring field is 0.2 T.}
\end{figure}

\begin{table}[t]
\caption{\label{tab:table2}Latent heats and critical temperatures of
first order ferromagnetic or antiferromagnetic transitions in some
rare earth elements, and rare earth based compounds.}
\begin{ruledtabular}
\begin{tabular}{cccc}
 Compound & T$_C$(K), T$_N$(K) & $L (J/mole RE)$ & Ref.\\
\hline
Dy & $91 (FM)$ & 39 & ~\cite{Jayasuriya}\\
Er & $19 (FM)$ & 24 & ~\cite{Durfee}\\
Er$_{0.4}$Ho$_{0.6}$Rh$_4$B$_4$ & $1.9 (FM)$ & $74.5$ & ~\cite{lachal}\\
Sm$_2$IrIn$_8$  & $14 (AFM)$ & $5$ & ~\cite{Pagliuso}\\
Sm$_{0.55}$Sr$_{0.45}$MnO$_3$ & $130 (FM)$ & $180$ & ~\cite{Chernyshov}\\
Yb$_3$Pt$_4$ & $2.4 (AFM)$ & $0.09$ & this work \\
\end{tabular}
\end{ruledtabular}
\end{table}
Measurements of the ac magnetic susceptibility $\chi'$ show that
Yb$_{3}$Pt$_{4}$ likely orders antiferromagnetically in zero field.
As shown in Fig.\ref{f6}a, the real part of $\chi'$ has a distinct
cusp, corresponding to a discontinuity in the temperature derivative
d$\chi'$/dT at 2.4 K, the temperature where the ordering transition
is observed in the heat capacity, C$_{M}$.
As indicated in Fig.\ref{f6}b, the antiferromagnetic transition is
revealed as a marked slope change at 2.4 K for the dc
susceptibilities $\chi_{\perp}^{-1}$ and $\chi_{\parallel}^{-1}$,
showing that the antiferromagnetic order develops directly from a
simple paramagnetic state without appreciable critical fluctuations
or substantial magnetic anisotropy. Fig. 6b shows that the dominant
component of the susceptibility in the magnetically ordered state is
large and temperature independent, approaching a value of
$\chi_{\perp}(0)$=6.8$\times$10$^{-6}$m$^3$/mole Yb, and $\chi_{\parallel}(0)$=3.2$\times$10$^{-6}$m$^3$/mole Yb.
\begin{figure}[t]
\begin{center}
\includegraphics[width=7cm]{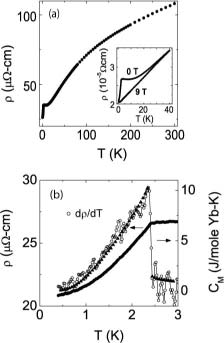}
\end{center}
\caption{\label{f7} (a) The temperature dependence of the electrical
resistivity $\rho$(T) of Yb$_{3}$Pt$_{4}$. Inset shows an expanded
view of the low temperature resistivity, in zero field and in a
magnetic field of 9 T applied along the c-axis. b) A comparison of
the temperature dependences of C$_{M}$(filled triangles), the
electrical resistivity $\rho$ (filled circles), and the temperature
derivative of the resistivity $\partial\rho/\partial$T (open
circles) near the antiferromagnetic transition temperature,
T$_{N}$=2.4 K.}
\end{figure}

The temperature dependence of the electrical resistivity $\rho$(T)
demonstrated in Fig. \ref{f7}a indicates that Yb$_{3}$Pt$_{4}$ is a
good metal. $\rho$(T) decreases from the room temperature value of
108 $\mu\Omega$-cm to a residual value of $\sim$ 21 $\mu\Omega$-cm
at 0.4 K.  $\rho$(T) has a broad maximum near 2.4 K, but drops by
$\sim$ 25$\%$ with the onset of magnetic order. As is evident from
Fig. \ref{f7}b, the temperature dependences of the heat capacity and
the temperature derivative of the resistivity
$\partial\rho/\partial$T are nearly identical, indicating that the
resistivity is controlled both above and below T$_{N}$ by
fluctuations in the magnetization.~\cite{fisher1976} Application of
a 9 T field (Fig. \ref{f7}a inset) completely suppresses this spin
disorder scattering, and renders the underlying resistivity linear
in temperature from 0.4 K - 40 K.

\begin{figure}[t]
\begin{center}
\includegraphics[width=7cm]{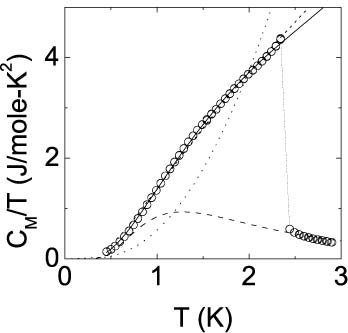}
\end{center}
\caption{\label{f8} The temperature dependence of the electronic and
magnetic part of the specific heat C$_{M}$ divided by temperature
(open circles) is fit to a the expression for a ferromagnetic spin
wave with a gap $\Delta$ (short-dashed line),
C$_M$/T=AT$^{0.5}$e$^{-\delta/T}$, where A=3.3 J/mole-Yb-K$^{5/2}$,
and $\delta$=$\Delta$/k$_B$=1.2 K. The dotted line is a fit to an
antiferromagnetic spin wave, C$_M$/T=AT$^{2}$e$^{-\delta/T}$,
assuming a 1 K gap. In both cases, the coefficient of the electronic
part of the specific heat, $\gamma$ is found to be negligible. The
best fit to the data in the ordered phase is by
C$_M$/T=AT$^0$e$^{-\delta/T}$ (solid line). The dashed line is the expected
Schottky anomaly if the doublet ground state is split.}
\end{figure}

Our measurements indicate that the antiferromagnetically ordered
state is a Fermi liquid, albeit one with unusual properties. We have
plotted the temperature dependence of C/T in Fig. \ref{f8}, along
with several candidate fits. It is clear that there is little evidence for a quasiparticle
contribution to the heat capacity $\gamma$=C/T. Spin wave expressions for C
are compared to the data in Fig. 8. Despite the bulk antiferromagnetic
order observed in the magnetic susceptibility measurements, C/T is
poorly described by C/T=T$^2$exp(-$\Delta$/k$_B$T), while reasonable agreement is
found for the expression corresponding to ferromagnetic spin waves
C/T=T$^{0.5}$exp(-$\Delta$/k$_B$T)with an anisotropy gap $\Delta$=1.2 K. This
may suggest that magnetic order in Yb$_3$Pt$_4$ results from the competition
between both ferromagnetic and antiferromagnetic interactions.  We note,
however, that the gradual increase in C/T above T$_N$ and the T$^2$ behavior
found at the lowest temperatures may be consistent with an underlying
Schottky peak, implying that the lowest lying Yb doublet may be split
by $\sim$1.6 K, perhaps by the exchange
interaction. Further experiments, such as measurements of the heat
capacity in field, would be needed to test this possibility. In either
case, the vanishingly small value of C/T found at the lowest
temperatures suggests that the magnetic excitations which lead to the
heat capacity in the ordered state are gapped. Seemingly, this would
rule out the possibility that the ordered state is a Fermi Liquid. 

Given this conclusion, it is surprising that the electrical
resistivity gives unambiguous evidence for Fermi liquid behavior in
the antiferromagnetically ordered state. Fig. 9 shows that
$\rho$=$\rho_{o}$+AT$^{2}$ for temperatures smaller than the
magnetic ordering temperature T$_{N}$=2.4 K, with $\rho_{o}$=21
$\mu\Omega$-cm, and A= 1.69$\times$10$^{-6}$ $\Omega$-cm/K$^{2}$. This
result shows that a substantial portion of the resistivity in the
antiferromagnetic state results from quasiparticle scattering in a
Fermi liquid, indicating that the relatively large resistivity does
not result from strong impurity scattering but is an intrinsic
feature of Yb$_{3}$Pt$_{4}$. The magnitude A of the quadratic
resistivity is very large, comparable to values found in heavy
electron systems such as
UPt$_{3}$.\cite{Visser,kadowaki1986,tsujii2003} Since the heat
capacity measurements rule out the presence of large numbers of
massive quasiparticles, we conclude that the large Fermi liquid
resistivity instead implies that there is a significantly reduced
number of quasiparticles in Yb$_{3}$Pt$_{4}$. We can roughly
estimate this number by assuming that the scattering rate per
quasiparticle is the same in Yb$_{3}$Pt$_{4}$ as in other Yb-based
heavy electron systems, i.e. that it has the same Kadowaki-Woods
ratio A/$\gamma^2$=1 $\mu\Omega$-cm
mol$^{2}$K$^{2}$/J$^{2}$.~\cite{kadowaki1986,
tsujii2003,kontani2004} Since our heat capacity measurements place
an upper bound on $\gamma$ of 10 mJ/mol-K$^{2}$, this implies that
the number of quasiparticles per Yb is approximately 10$^{-3}$.
Electronic structure calculations are needed to confirm this
hypothesis.

The antiferromagnetically ordered state is characterized by an
extremely large magnetic susceptibility. Fig. 6a shows that the ac
susceptibility in the ordered phase has a temperature independent
component which extrapolates to 6.8$\times$10$^{-6}$m$^{3}$/mole-Yb.
The temperature independence of this susceptibility suggests that it
may be the Pauli susceptibility of a Fermi liquid. However, we note
that this is a very large value, similar to those found in heavy
electron systems, both in absolute terms and especially when
weighted per quasiparticle. The Sommerfeld-Wilson ratio
R$_{W}$=$\pi^{2}$k$_{B}^{2}$$\chi_{o}$/$\mu_{eff}^{2}\gamma$ is a
useful figure of merit for judging the strength of ferromagnetic
correlations within a Fermi Liquid. Most heavy electron compounds
have R$_{W}$ somewhat larger than the Kondo value of
2.~\cite{hewson,fisk1987} Systems with ferromagnetic enhancement can
have significantly larger values of R$_{W}$,  such as Pd, where
R$_{W}$=6-8,~\cite{mueller1970} YbAgGe, where
R$_{W}$=2,~\cite{tokiwa2006} YbRh$_{2}$Si$_{2}$ where
R$_{W}$=13,~\cite{gegenwart2005} and also Sr$_{3}$Ru$_{2}$O$_{7}$
where R$_{W}$=10.~\cite{ikeda2000} The magnitude of the zero
temperature susceptibility in Yb$_{3}$Pt$_{4}$ is comparable to
those of the most enhanced heavy electron compounds,~\cite{fisk1987}
although the Sommerfeld coefficient $\gamma$ is several orders of
magnitude smaller, implying an unreasonably large value for
R$_{W}\sim$3600. We believe that a more likely scenario is that,
like the heat capacity, the temperature dependence of the ac
susceptibility is dominated by fluctuations of presumably localized
Yb moments and not the Pauli susceptibility of the few
quasiparticles which provide the electrical resistivity.

Finally, it is worth noting that measurements have been performed on
polycrystalline samples of the isostructural compound
Yb$_{3}$Pd$_{4}$ which have some similarity to the results reported
here for Yb$_{3}$Pt$_{4}$. Magnetic order is found in
Yb$_{3}$Pd$_{4}$ at 3 K,~\cite{politt1985} and strong critical
scattering is observed in inelastic neutron scattering measurements.
~\cite{walter} Magnetic susceptibility and x-ray absorption
measurements~\cite{politt1985} find that this compound is - unlike
Yb$_{3}$Pt$_{4}$ - significantly mixed valent. It might be
interesting to explore this system more thoroughly, in particularly
to see whether this mixed valent character could be suppressed by
pressure, providing a direct connection to Yb$_{3}$Pt$_{4}$.

\begin{figure}[t]
\begin{center}
\includegraphics[width=7cm]{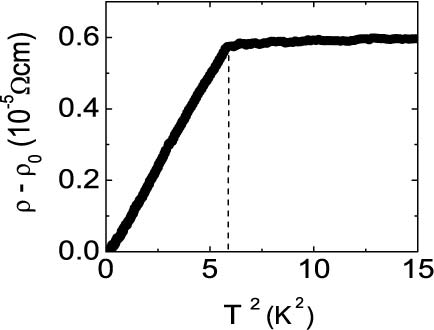}
\end{center}
\caption{\label{f9} The temperature dependent part of the electrical
resistivity $\Delta\rho$=$\rho$-$\rho_{o}$ is quadratic in
temperature below T$_N$=2.4 K (dashed line).}
\end{figure}

\section{conclusion}

The picture which emerges from our measurements on the new compound
Yb$_{3}$Pt$_{4}$ is that this is system which orders overall
antiferromagnetically, with very little magnetic anisotropy. The
transition is weakly first order, and there is little indication of
critical fluctuations in either the magnetic susceptibility or the
heat capacity. The large magnitude of the low temperature magnetic
susceptibility, the observation of a Fisher-Langer relation between
the heat capacity and the temperature derivative of the resistivity,
as well as the anisotropy gap found in the ordered phase heat
capacity all suggest that the magnetic structure has a net
ferromagnetic component as well. A possible explanation might be
that the ground state of Yb$_{3}$Pt$_{4}$ involves a long-wavelength
ferromagnetic modulation of the antiferromagnetic state similar to
that found in the helical ferromagnet MnSi.~\cite{pfleiderer2004}
The temperature dependence of the heat capacity and the magnitude of
the magnetic susceptibility together indicate that the localized
moments continue to be an important part of the physics of the
ordered phase.

The most remarkable of our findings is that magnetic order enables
the formation of a Fermi liquid state in Yb$_{3}$Pt$_{4}$.
Antiferromagnetic order emerges from a paramagnetic state involving
fluctuations of individual Yb moments, and there is little evidence
that the apparently localized moments are strongly coupled to
conduction electrons via the Kondo effect, as they are in quantum
critical systems such as the heavy electron compounds. The
electrical resistivity is large for this class of intermetallics,
both in the paramagnetic and ordered phases. However, the
substantial loss of spin-disorder scattering associated with the
magnetic transition indicates that low carrier density and not
strong impurity scattering is likely the primary source of this high
but decidedly metallic resistivity. Given this apparent weak
coupling between the magnetic moments and the conduction electrons,
it is somewhat surprising to find that the resistivity in the
ordered state is that of a Fermi liquid: quadratic with temperature
$\rho$(T)=$\rho_{0}$+AT$^{2}$, and with a coefficient A as large as
those found in heavy electron compounds. The absence of a large
quasiparticle contribution to the heat capacity subsequently implies
that the Fermi surface in the ordered phase has a small volume,
perhaps consisting of isolated pockets.

The central issues surrounding the suitability of an ordering
material as an eventual host for quantum criticality are the
classification of the phase transition, with respect to order and
mechanism, and the ways in which the underlying electronic structure
is affected by the onset of magnetic order. While we know that the
magnetic ordering phase transition is dominantly antiferromagnetic
and weakly first order, the overall classification of the phase
transition in Yb$_{3}$Pt$_{4}$ into local moment type or Fermi
surface instability is not wholly settled by our experiments. On the
one hand, spin wave expressions provide a reasonable description of
the ordered phase heat capacity, however the magnitude of the
magnetic susceptibility is much too large to be conveniently
assigned to itinerant Fermi surface states. In the local moment
ordering scenario, superzone formation can potentially gap the Fermi
surface, although the low crystal symmetry of Yb$_{3}$Pt$_{4}$
suggests that this gapping would be at best partial. The large value
of the quasiparticle resistivity is consistent with the formation of
a Fermi liquid comprised of the few states at the Fermi energy which
survive this gapping. While it would be quite important to know
whether these quasiparticles have substantial mass enhancement, a
definitive determination would require knowing the quasiparticle
contributions to the magnetic susceptibility and heat capacity,
which are apparently masked by the larger local moment
contributions. However, the onset of Fermi liquid behavior exactly
at the Neel temperature suggests that antiferromagnetic order
involves a more profound modification to the Fermi surface and its
excitations than is afforded by simple superzone formation. We
cannot rule out the possibility that a Fermi surface instability
plays a role in driving magnetic order in Yb$_{3}$Pt$_{4}$. More
information about the magnetic structure and dynamics in the ordered
state, as well as direct determinations of the Fermi surface above
and below the transition are needed to provide a definitive
resolution of this issue.

\begin{acknowledgments}
The authors acknowledge useful conversations with C. Varma, Q. Si,
J. W. Allen, P. Coleman, J. Kampf and P. Stephens. We are grateful
to J. Chan and J. Millican for crystallographic consultations.
Electron microscopy was carried out at the University of Michigan
Electron Microbeam Analytical Laboratory (EMAL). Work at the
University of Michigan and at Stony Brook University was supported
by the National Science Foundation under grant NSF-DMR-0405961.
\end{acknowledgments}


\begin{thebibliography}{99} 
\bibitem{doniach} S. Doniach, in {\it Valence Instabilities and
Related Narrow Band Phenomena}, edited by R. D. Parks (Plenum, New
York, 1977), p. 169.
\bibitem{stewart2001}G. R. Stewart, Rev. Mod. Phys. $\bf{73}$, 797 (2001).
\bibitem{gegenwart2002}P. Gegenwart, J. Custers, C. Geibel, K.
Neumaier, T. Tayama, K. Tenya, O. Trovarelli, and F. Steglich, Phys.
Rev. Lett. {\bf 89}, 056402 (2002).
\bibitem{grigera2004}S. A. Grigera, P. Gegenwart, R. A. Borzi, F. Weickert,
A. J. Schofeld, R. S. Perry, T. Tayama, T. Sakakibaram Y. Maeno, A.
G. Green, and A. P. Mackenzie, Science $\bf{306}$, 1154 (2004).
\bibitem{vonlohneysen1996}H. von Lohneysen, J. Phys.: Cond. Matt. $\bf{8}$, 9689 (1996).
\bibitem{fisk1989} Z. Fisk and J. P. Remeika, in {\it Handbook on the Physics and Chemistry of the Rare Earths, Vol. 12}, edited by K. A. Gschneidner and L. Eyring
(Elsevier Science Publishers B. V., 1989), p. 53.
\bibitem{palenzona1977}A. Palenzona, J. Less-Common Metals
$\bf{53}$, 133 (1977).
\bibitem{politt1985} B. Politt, D. Durkop and P. Weidner, J. Magn. Magn. Mat.,
$\bf{47\&48}$, 583-585 (1985).
\bibitem{walter} U. Walter, D. Wohlleben, Phys. Rev. B $\bf{35}$, 3576 (1987).
\bibitem{sheldrick} G. M. Sheldrick, SHELXL-97, Program for Crystal Structure Refinement, University
of G\"ottingen, Germany, 1997.
\bibitem{sales2000} B. C. Sales, D. G. Mandrus, and B. C. Chakoumakos, in {\it Recent Trends in Thermoelectric Materials Research II}, Academic Press, San Diego,
California, P. 1 (2000).
\bibitem{sales1999} B. C. Sales, B. C. Chakoumakos, D. Mandrus, and J. W. Sharp, J. Solid State Chem. $\bf{146}$, 528-532 (1999).
\bibitem{Jayasuriya} K. D. Jayasuriya, S. J. Campbell, A. M. Stewart, Phys. Rev. B {\bf 31}, No. 9,
6032 (1985).
\bibitem{Durfee} C. S. Durfee and C. P. Flynn, Phys. Rev. Lett. {\bf 87}, No. 5,
057202-1 (2001).
\bibitem{lachal} B. Lachal, M. Ishikawa, A. Junod, and J. Muller, J. Low Temp. Phys. {\bf 46}, No. 5/6
467 (1982).
\bibitem{Pagliuso} P. G. Pagliuso, J. D. Thompson, M. F. Hundley, J. L. Sarrao, and Z. Fisk,
Phys. Rev. B {\bf 63}, 054426 (2001).
\bibitem{Chernyshov} A. S. Chernyshov, M. I. Ilyn, A. M. Tishin, O. Yu. Gorbenko, V. A. Amelichev,
S. N. Mudretsova, A. F. Mairova, and Y. I. Spichkin , Cryocoolers
{\bf 13}, 381 (2004).
\bibitem{fisher1976} M. E. Fisher and J. S. Langer, Phys. Rev. Lett. $\bf{20}$, 665 (1968); S. Alexander, J. S. Helman, and I. Balberg,
Phys. Rev. B {\bf 13}, 304 (1976).
\bibitem{ramirez2001}A. P. Ramirez, in {\it Handbook of Magnetic
Materials, Vol. 13}, edited by K. H. J. Buschow (Elsevier, Amsterdam
2001) p. 423
\bibitem{janssen2008}Y. Janssen, M. S. Kim, M. C. Bennett, Q. Ying,
J. W. Lynn, and M. C. Aronson (unpublished).
\bibitem{millis1993}A. J. Millis, Phys. Rev. B $\bf{48}$, 7183
(1993).
\bibitem{Visser} A. de Visser, J. J. M. Franse, and A. Menovsky, J. Magn. Magn.
Mater. {\bf 43}, 43 (1984).
\bibitem{kadowaki1986}K. Kadowaki and S. B. Woods, Solid State
Commun. $\bf{58}$, 507 (1986).
\bibitem{tsujii2003}N. Tsujii, K. Yoshimura, and K. Kosuge, J. Phys.
Cond. Matt. $\bf{15}$, 1993 (2003).
\bibitem{kontani2004}H. Kontani, J. Phys. Soc. Japan $\bf{73}$, 515
(2004).
\bibitem{hewson}A. C. Hewson, {\it The Kondo Problem to Heavy
Fermions, \rm} (Cambridge University Press 1993) p. 15
\bibitem{fisk1987}Z. Fisk, H. R. Ott, and G. Aeppli, Japanese
Journal of Appl. Phys. $\bf{26},Suppl. 26-3$, 1882 (1987).
\bibitem{mueller1970}F. M. Mueller, A. J. Freeman, J. O. Dimmock,
and J. M. Furdyna, Phys. Rev. B $\bf{1}$, 4617 (1970).
\bibitem{tokiwa2006}Y. Tokiwa, A. Pikul, P. Gegenwart, F. Steglich,
S. L. Bud'ko, and P. C. Canfield, Phys. Rev. B $\bf{73}$, 094435
(2006).
\bibitem{gegenwart2005}P. Gegenwart, J. Custers, Y. Tokiwa, C.
Geibel, and F. Steglich, Phys. Rev. Lett. $\bf{94}$, 076402 (2005).
\bibitem{ikeda2000}S. -I. Ikeda, Y. Maeno, S. Nakatsuji, M. Kosaka,
and Y. Uwatoko, Phys. Rev. B $\bf{62}$, R6089 (2000).
\bibitem{pfleiderer2004} C. Pfleiderer, D. Reznik, L. Pintschorius, H. V. Lohneysen, M. Garst and A. Rosch,
Nature {\bf 427}, 227 (2004).
\end{thebibliography}
\end{document}